Electronic structure and optical band gap of $CoFe_2O_4$ thin films

A. V. Ravindra and P. Padhan

Department of Physics, Indian Institute of Technology Madras,

Chennai – 600036, India.

W. Prellier

Laboratoire CRISMAT, CNRS UMR 6508, ENSICAEN,

6 Bd du Marechal Juin, F-14050 Caen Cedex, FRANCE.

Abstract

Electronic structure and optical band gap of $CoFe_2O_4$ thin films grown on (001) oriented $LaAlO_3$ have been investigated. Surprisingly, these films show additional Raman modes at room temperature as compared to a bulk spinel structure. The splitting of Raman modes is explained by considering the short-range ordering of Co and Fe cations in octahedral site of spinel structure. In addition, an expansion of band-gap is observed with the reduction of film thickness, which is explained by the quantum size effect and misfit dislocation. Such results provide interesting insights for the growth of spinel phases.

The arrangement of transition metal ions in oxide magnetic materials plays a crucial role on its structure as well as on physical properties[1-3]. In particular the arrangement of transition metal ions in tetrahedral and/or octahedral sites of $AB_2O_4$ spinel structure is very appealing. Very recently Ivanov et al. have addressed this issue using polarization Raman spectroscopy since the number, the frequencies, and the polarization selection rules of the Raman-active vibrational modes are highly sensitive to the atomic short-range order[1]. Also, using first principle density functional theory Daniel et al.[4] have obtained significant energy differences between various cation arrangements with the same degree of inversion, providing further evidence for short range B site order. Among the various ferrites of spinel structure, the cobalt ferrite ($CoFe_2O_4$ ($CFO$)) exhibits several interesting electronic properties such as high Curie temperature, relatively high saturation magnetization, diffusion anisotropy and good chemical stability. The $CFO$ has been used to fabricate heterostructures for spin filtering[5], resistive switching[6] and magnetoelectric coupling between ferromagnetic and ferroelectric thin film layers [7]. Since the band gap or the barrier height of the ferromagnetic (ferrimagnetic) insulator is considered to be spin-dependent, quite high spin polarization of the current is indeed expected due to spin-filter effect. It has also been recognized that size effect and residual strain play an important role in controlling the coercivity and other magnetic properties [8, 9]. Moreover, ferrites offer the advantage of having a band gap capable of absorbing visible light, as well as, the spinel crystal structure, which enhances efficiency due to the available extra catalytic sites by virtue of the crystal lattice[10]. Therefore, synthesizing highly single crystalline $CFO$ thin films and establishing a comprehensive understanding on the optical properties, electronic structures, and coupling effects of these ferrites is essential not only to improve thin film quality but also to develop potential applications. In this letter, we present the study of optical properties of $CFO$ thin films grown on (001) oriented $LaAlO_3 (LAO)$ substrates. Such results particularly highlight the role of thickness on the optical properties of $CFO$ thin films.

A series of $CFO$ thin films with different thicknesses were grown on (001)-oriented $LAO$ substrates using pulsed laser deposition technique. Briefly, a KrF excimer laser $(248\,nm, 200\,mJ)$ beam was focused onto a $2\,cm$ diameter rotating target of stoichiometric composition of $CFO$, leading to an energy density of $2\,J/cm^2$. Highly dense ablation targets were prepared through the standard ceramic route using $99.9\,\%$ pure $CoCO_3$ and $FeCO_3$ as starting materials. Thin film deposition was carried out at $700\,°C$ in $25\,mTorr$ pressure of oxygen. The typical deposition rate for $CFO$ was calibrated to $\sim 0.058\,Å/pulse$. After the completion of thin film deposition, the chamber was filled with oxygen of $300\,Torr$, and the sample was cooled down to room temperature at the rate of $15\,°C/min$.

The crystallinity and epitaxy of the films were examined in a $\theta - 2\theta$ x-ray diffractometer. The film thicknesses were calculated using established deposition rate of $CFO$ obtained from the angular positions of interference fringes of low angle $\theta - 2\theta$ x-ray profiles. The film stoichiometry was checked with electron probe microanalysis on typical samples and was found to be in agreement with the expected composition. The Raman spectra were recorded on a Jobin-Yvon LabRAM HR800UV spectrometer instrument equipped with highly efficient Thermo-Electrically cooled charge coupled device (CCD). The spectra were taken at room temperature in the backscattering configuration using $633\,nm$ emission line of an He-Ne laser with lower than $2\,mW$ laser power on the sample surface. The absorbance of $CFO$ at room temperature was measured using a dual-beam JASCO V-570 UV/Vis/NIR spectrophotometer in the wavelength range of $200 - 800\,nm$, with a spectral resolution of $1\,nm$.

The lattice parameter of bulk $LAO$ and $CFO$ is 3.792 Å and 8.39 Å, respectively. Thus, $LAO$ provides in-plane compressive stress for the epitaxial growth of $CFO$ with lattice mismatch +10.62 %. According to Frank and Van der Merwe growth mode the lattice of the film can be suppressed by the lattice of the substrate through elastic strain and epitaxial growth can continue in a layer-by-layer manner if the lattice mismatch is less than 7 %[11]. However, even with such a high value of lattice mismatch (+10.62 %), the $CFO$ thin film stabilizes with cubic spinel phase. The $\theta - 2\theta$ x-ray scans of thin films of $CFO$ show indeed only the $(00l)$ Bragg's reflections of the substrate and $CFO$, indicating the epitaxial nature of cubic spinel phase of $CFO$ thin films. Fig. 1 shows the x-ray measurements of $CFO$ thin films grown on $LAO$. The average out-of-plane lattice parameter obtained from the $(00l)$ diffraction peak position is 8.121, 8.357 and 8.362 Å for 58, 363 $and$ 581 Å thick film of $CFO$, respectively. Thus, the misfit-strain occurring in the 58, 363 $and$ 581 Å thick films of $CFO$ grown on $LAO$ is calculated to be $\sim$ 3.20, $\sim$ 0.39 and $\sim$ 0.33 % respectively. Similar value of strain obtained for $\sim$ 363 $and$ $\sim$ 581 Å thick $CFO$ films suggests that the effect of substrate induced stress on these two films is not significant.

Figure 2 shows the low angle $\theta - 2\theta$ x-ray diffraction spectrum (open circle curve) of a $CFO$ thin film. The red curve (see Fig. 2) represents the fit using Philips WINGIXA™ software to the measured low angle $\theta - 2\theta$ x-ray diffraction scans. The Presence of Kiessig fringes confirms that the surfaces and interfaces of the thin film are extremely smooth. Clearly, the relative intensity and the angular position of the Kiessig fringes of the x-ray profile are in good agreement with the simulated one. From the fit the values of the density, surface roughness and thickness (t) of $CFO$ are found to be $\sim 5.12 \ gm/cm^3$, $\sim$5.95 Å and $\sim$ 363 Å, respectively. The thickness of thicker films is also calculated from the positions of the extrema (θ) of the specularly reflected x-ray beam. The positions of the extrema are given by $sin^2\theta = 2\delta + \left(\frac{n\lambda}{2t}\right)^2$, where '$\delta$' is defined by a quantum mechanical dispersion relation, '$n$' is the order of the extremum and '$\lambda$' is the wavelength[12]. The linear variation of $sin^2\theta$ with $n^2$ of ~ 363 Å thick film of $CFO$ on $LAO$ is shown in the inset of Fig. 2. The thickness calculated from the slope of $sin^2\theta$ vs. $n^2$ line of ~ 363 Å and 581 Å thick $CFO$ films is ~ 361.8 Å and ~ 581.82 Å, respectively. The thicknesses of these two thin films calculated from both the methods are comparable. However, the number of Kiessig fringes i.e. '$n$' observed for ~ 58 Å thick $CFO$ is not sufficient to plot $sin^2\theta$ vs. $n^2$ graph.

Figure 3 shows the room temperature Raman spectra of substrate and the series of $CFO$ films. Note that a single sharp peak is located at about $487.45\ cm^{-1}$ in the Raman spectrum of the substrate which corresponds to the Raman mode of the single crystalline $LAO$. As ~ 58 Å thick $CFO$ is grown on the substrate, additional peaks are observed in its Raman spectra. This ~ 58 Å thick film shows apparently noticeable peaks at about 465 and $704\ cm^{-1}$ and weak peaks at 342, 584 and $671\ cm^{-1}$. The weak peaks observed in the Raman spectrum of ~ 58 Å thick film are attributed to the microstructure evolution, which is correlated with its vertical size effect. However, as the thickness of the $CFO$ film increases, the remarkable Raman spectra peaks become sharper, while the weak peaks become pronounced. In fact, except for small differences in phonon line parameters, the spectra and their variation as a function of the film thickness closely resemble to those of bulk $CFO$. According to the factor group analysis these phonon lines correspond to the optical active Raman modes ($A_{1g} + E_g + 3T_{2g}$) of cubic spinel with $Fd3m - O_h^7$ space group [13]. For cubic spinel, the $A_{1g}$ mode is assigned to the motion of oxygen in the $FeO_4$ tetrahedra due to the symmetric stretching of oxygen atom with respect to metal ion in tetrahedral void. The low frequency phonon modes $E_g$ and $T_{2g}$ correspond to the symmetric and anti–symmetric bending of oxygen($O$) atoms in $M - O$ ($M = Fe$ or $Co$) bond at $MO_6$ octahedral void respectively[14]. The Raman active mode frequencies of all thin films of $CFO$ (see Table – I) are assigned based on the $D_{3d}$ perturbed octahedral symmetry by which one of the $T_{2g}$ mode splits into two modes ($A_{1g}^*$ and $E_g^*$) as suggested by Graves *et. al.*[15]. Surprisingly, as the film thickness increases from $\sim 58$ Å to $\sim 581$ Å the number of experimentally observed Raman lines in the spectra of $CFO$ exceeds significantly the number expected for a normal spinel structure in bulk (see Table – I). It is remarkable, however, that these spectra are practically matching to those reported earlier for $FeFe_2O_4, NiFe_2O_4,$ and $CoFe_2O_4$ [15, 16]. The presence of extra Raman lines in the Raman spectra has in fact been observed for a wide variety of materials such as $ZrTiO_4$[17], $La_{0.5}Ca_{0.5}MnO_3$[18] and $NiFe_2O_4$[1]. The observed extra Raman lines in the Raman spectra of these materials has been explained in terms of the short range ordering. Therefore, the additional lines in the Raman spectra of $CFO$ are most likely related to a short-range order of $Fe$ and $Co$ cations. In addition to the observed extra Raman lines, the $CFO$ thin films also show most significant effect of softening of $T_{2g}$, $A_{1g}^*, E_g^*$ and $A_{1g}$ modes as the film thickness decreases [unlike the case of $NiFe_2O_4$ (ref. 2)]. In contrast, the shifts of $E_g$ Raman line with the thickness of $CFO$ film is not so clear due to the weak lines observed in the Raman spectrum of $\sim 58$ Å thick film.

The optical absorbance spectra in the higher energy region determine the band gap of the material, which is a crucial parameter for the barrier height in spin filter devices[5]. Thus, the optical absorbance spectra of $CFO$ thin films were recorded at room temperature. Fig. 4 shows the wave length dependence of the optical absorbance spectra of three $CFO$ thin films. From the absorbance spectra, it is seen that the absorbance of $\sim 58$ Å thick film at around 850 nm wavelength is negligibly small, but it increases gradually as the wavelength decreases up to 600 nm followed by a rapid increase of absorbance down to 360 nm wavelength. For the higher thickness films the absorbance spectra are qualitatively similar to that of the $\sim 58$ Å thick film, with a variation of absorption edge position. However, the position of the absorption edge differs significantly for $\sim 58$ Å thick $CFO$ film compared to the higher thickness films. A clear tendency in the variation of absorption edge is observed with the change of thickness of $CFO$ film. The absorption coefficient is extracted from the absorbance for each wavelength using Beer-Lambert law. The variation of absorption coefficient of $CFO$ thin film with wavelength is analyzed using a classical Tauc approach [19]. The $(\alpha h\nu)^2$ of different $CFO$ thin films with the variation of $h\nu$ is shown in Fig. 5. It can be seen that the plot varies linearly for all the films of $CFO$ in the region of strong absorption near the fundamental absorption edge. The linear variation of absorption coefficient of the $CFO$ at high frequencies indicates that these thin films have direct transitions across the energy band gap. The energy band gap was determined by extrapolating the linear part of $(\alpha h\nu)^2$ vs $h\nu$ curve to $h\nu = 0$. The band gap of $581, 363$ and $58$ Å thick films calculated from the linear fit is $2.505, 2.556$ and $2.615$ eV, respectively. These observed values of band gap of all the $CFO$ films are larger than the value $\sim 1.44\ eV$ of $1000$ Å thick film of $CFO$ grown on sapphire[20]. It is important to note that the energy band gap decreases as the film thickness increases. On the other hand, it can be seen from the inset of Fig. 1 that, with the increase of film thickness the magnitude of misfit strain in the film decreases significantly due to the relaxation. The strained film of $CFO$ is considered as a one-dimensional confined system. For one-dimensional confinement, the band gap expansion ($\Delta E_{gap}$) can be generally explained with reduction of film thickness[21] i.e. $\Delta E_{gap} \sim \frac{\hbar^2}{2mt^2}$, where 't' is the thickness of the layer and '$m$' is the effective carrier mass. However, the observed band gap expansion cannot be explained quantitatively due to the presence of high value of misfit strain for the thinner sample which can lead to misfit dislocation[22, 23].

In conclusion, the crystal structure, electronic structure and optical band-gap energies of a series of $CoFe_2O_4$ thin films grown on (001) oriented $LaAlO_3$ substrates using pulsed laser deposition techniques were studied. We found that the number of experimentally observed Raman lines in these spectra of $CFO$ exceeds significantly the expected number for a normal spinel structure. We explained these results by a short-range ordering of Fe and Co cations in octahedral site of spinel structure. These thin films also show expansion of band-gap with the reduction of film thickness, which is explained by the quantum size effect and misfit dislocation. These results will be useful for an understanding of spinel structure when deposited as a thin film.

Acknowledgments:

We greatly acknowledge Indo-French collaboration through the financial support of both the LAFICS and the IFPCAR/CEFIPRA (Project N°3908-1).

| Raman shift (Cm$^{-1}$) of $NiFe_2O_4$ powder (ref. 15) | Mode | Raman shift (Cm$^{-1}$) of $CoFe_2O_4$ with different thickness (Å) | | |
|---|---|---|---|---|
| | | 581 Å | 363 Å | 58 Å |
| 339 | $E_g$ | 334 | 338 | 342 |
| | | 452 | 448 | - |
| 490 | $T_{2g}$ | 469 | 467 | 465 |
| 579 | $A_{1g}^*$ | 578 | 572 | - |
| | | 594 | 593 | 584 |
| 666 | $E_g^*$ | 679 | 677 | 671 |
| 700 | $A_{1g}$ | 706 | 705 | 704 |

Table I : Raman shift of bulk spinel (i.e. polycrystalline powder of $NiFe_2O_4$ from ref. 15) and mode assignments for various $CoFe_2O_4$ films grown on $LaAlO_3$ substrates.

Figure captions:

Figure 1: The $\theta - 2\theta$ x-ray diffraction spectrum of $363$ Å thick $CoFe_2O_4$ film grown on (001) oriented $LaAlO_3$ substrate. The inset shows $\theta - 2\theta$ x-ray diffraction scans around the (004) Bragg's peak of $58, 363$ $and$ $581$ Å thick $CoFe_2O_4$ films.

Figure 2: The experimental and simulated (Philips WINGIXA™) low angle $\theta - 2\theta$ x-ray diffraction profiles of $363$ Å thick $CoFe_2O_4$ film grown on (001) oriented $LaAlO_3$ substrate. The inset shows the $sin^2\theta$ vs. $n^2$ plot of the $363$ Å thick $CoFe_2O_4$ film. The solid line in the inset is the fit.

Figure 3: Raman spectra of substrate and thin films of $CoFe_2O_4$ with different thickness grown on (001) oriented $LaAlO_3$.

Figure 4: Absorbance spectra of thin films of $CoFe_2O_4$ with different thickness grown on (001) oriented $LaAlO_3$ substrates.

Figure 5: plots of $(\alpha h\nu)^2$ vs. $h\nu$ of thin films of $CoFe_2O_4$ with different thickness grown on (001) oriented $LaAlO_3$ substrates. The plots also include a linear extrapolation of $(\alpha h\nu)^2$ to $h\nu = 0$ suggests a direct band gap.

Fig. 1

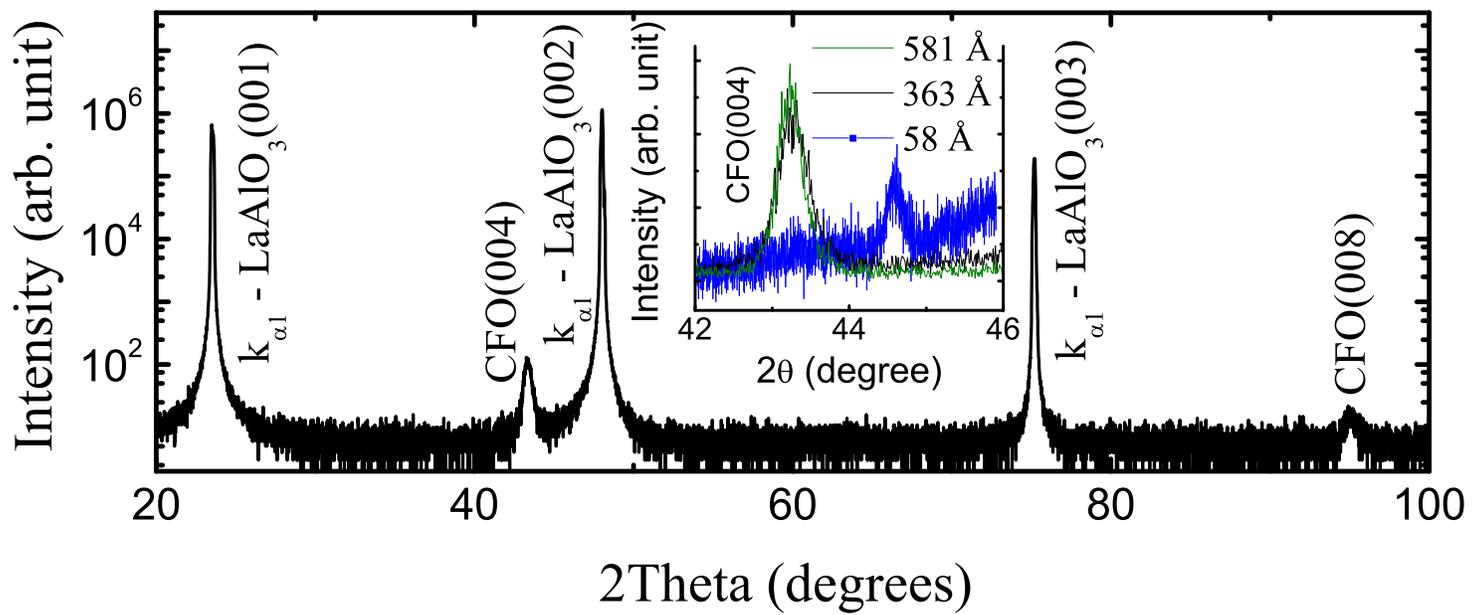

Fig. 2

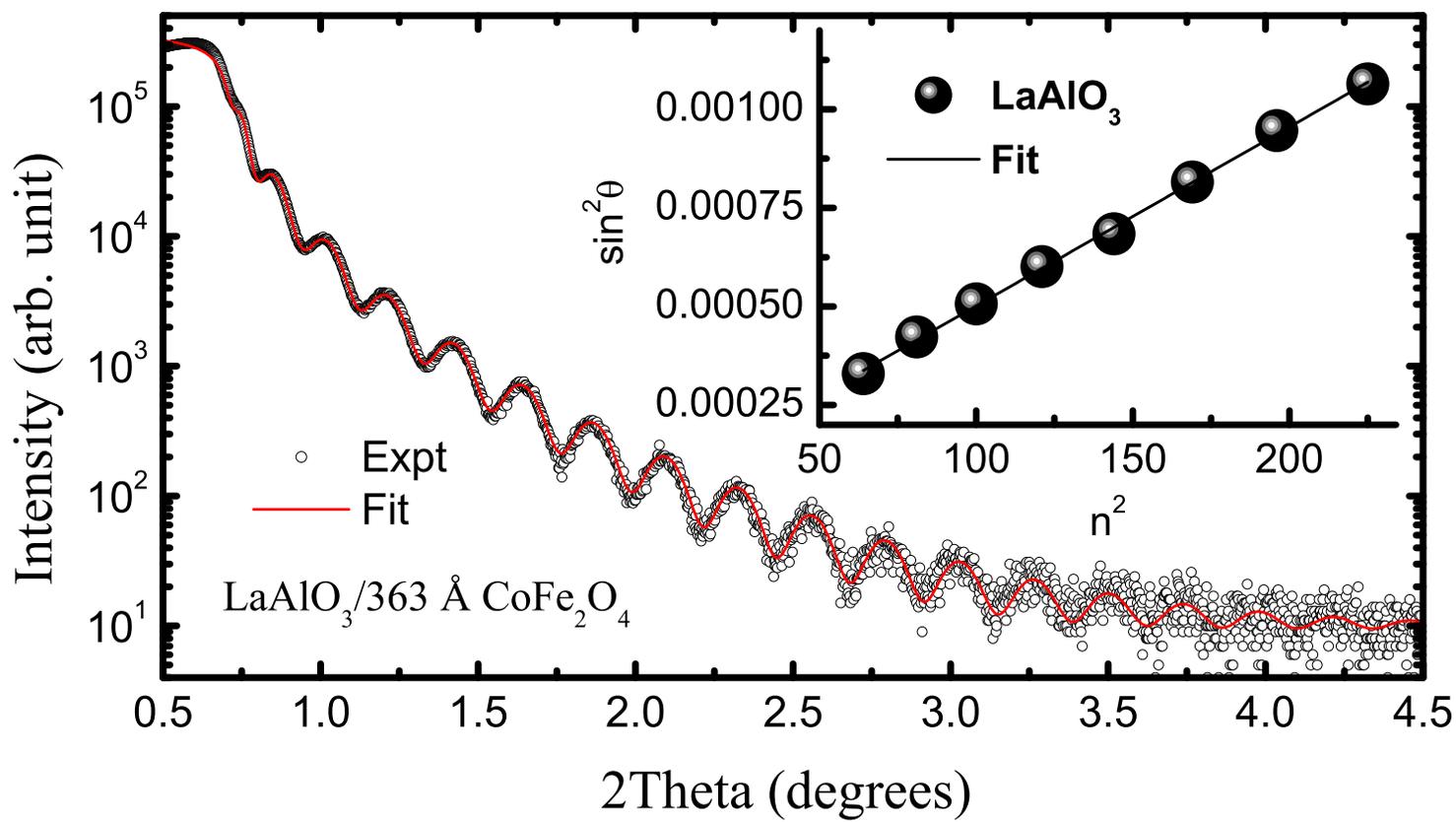

Fig. 3

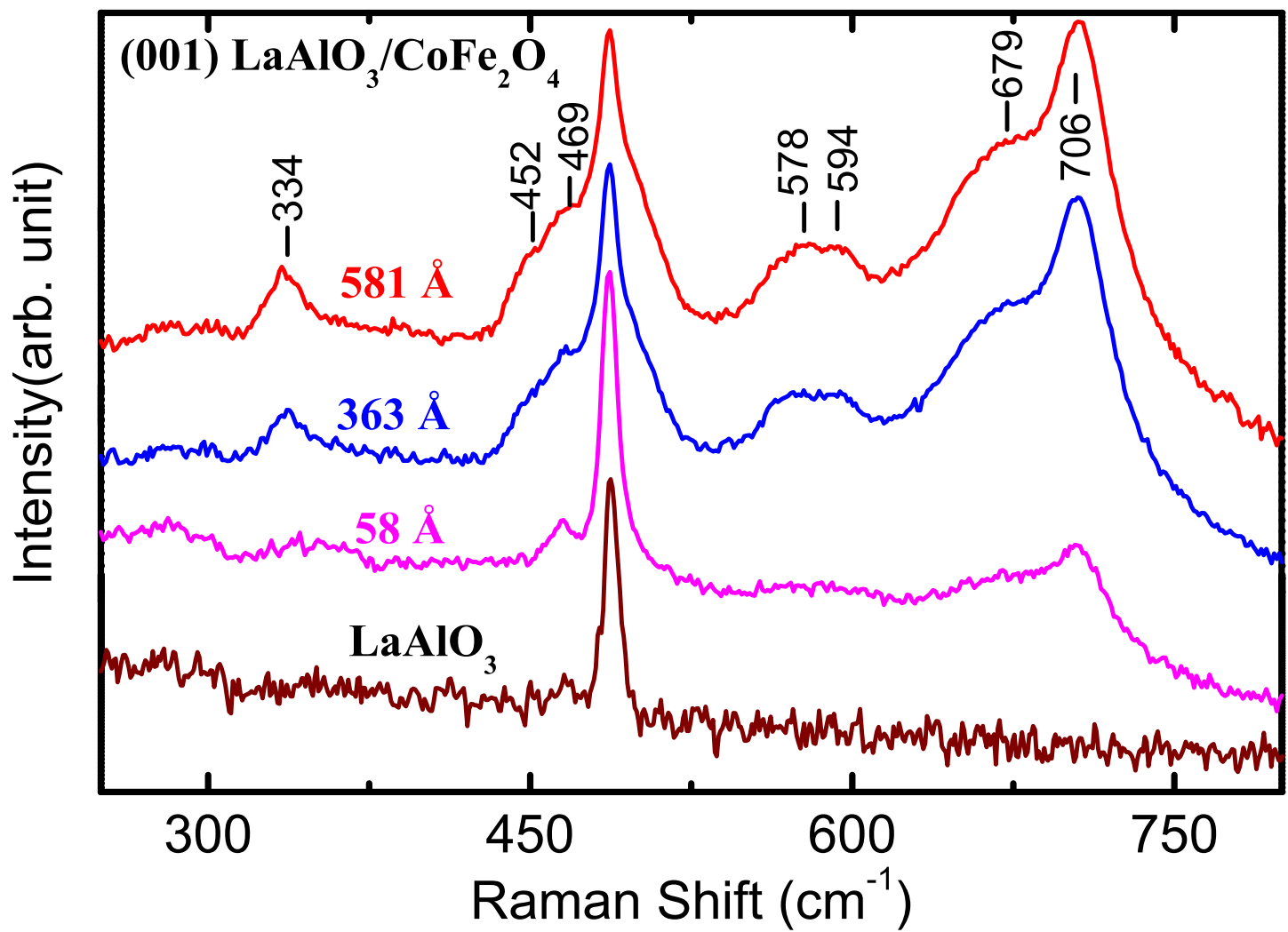

Fig. 4

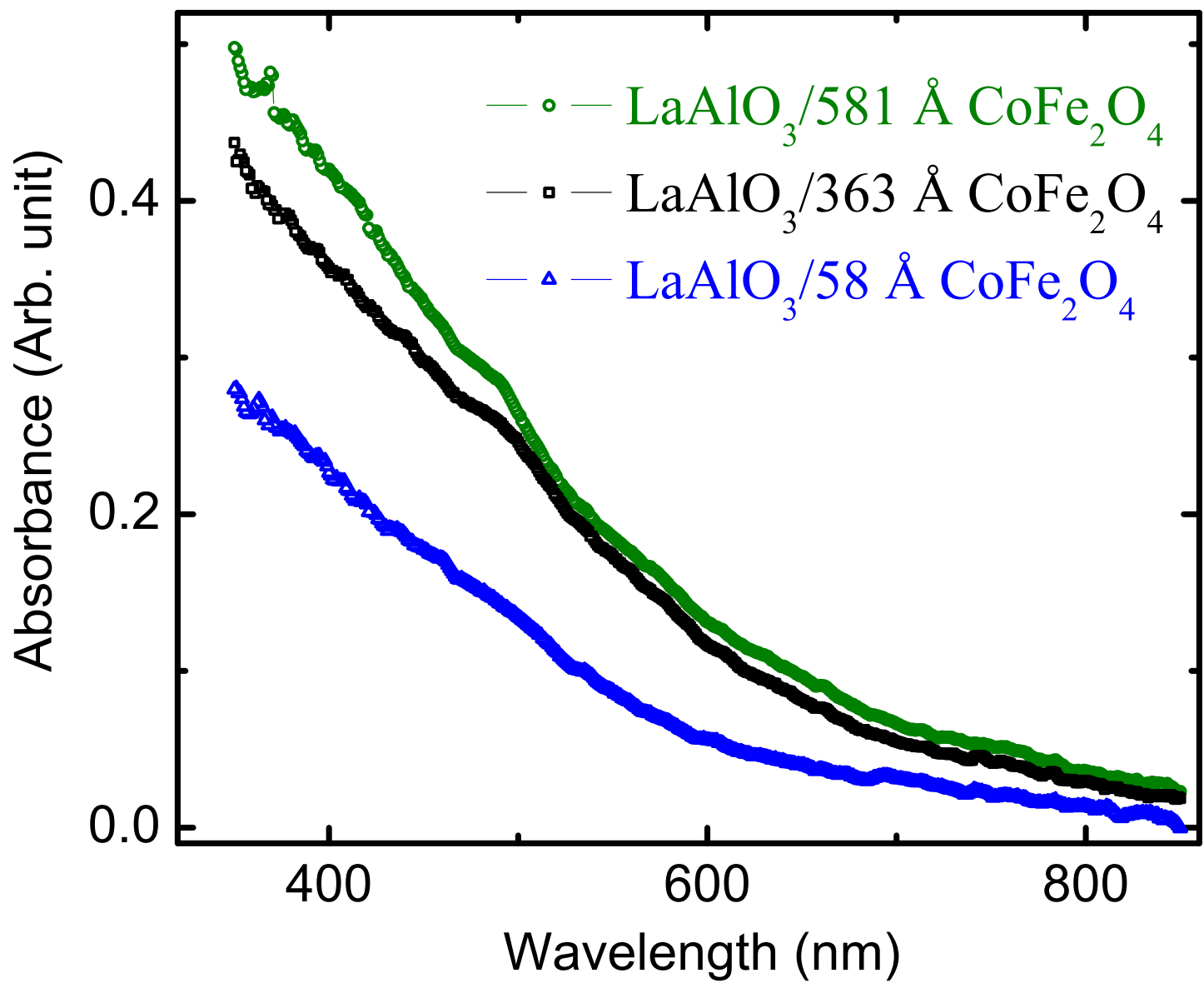

Fig. 5

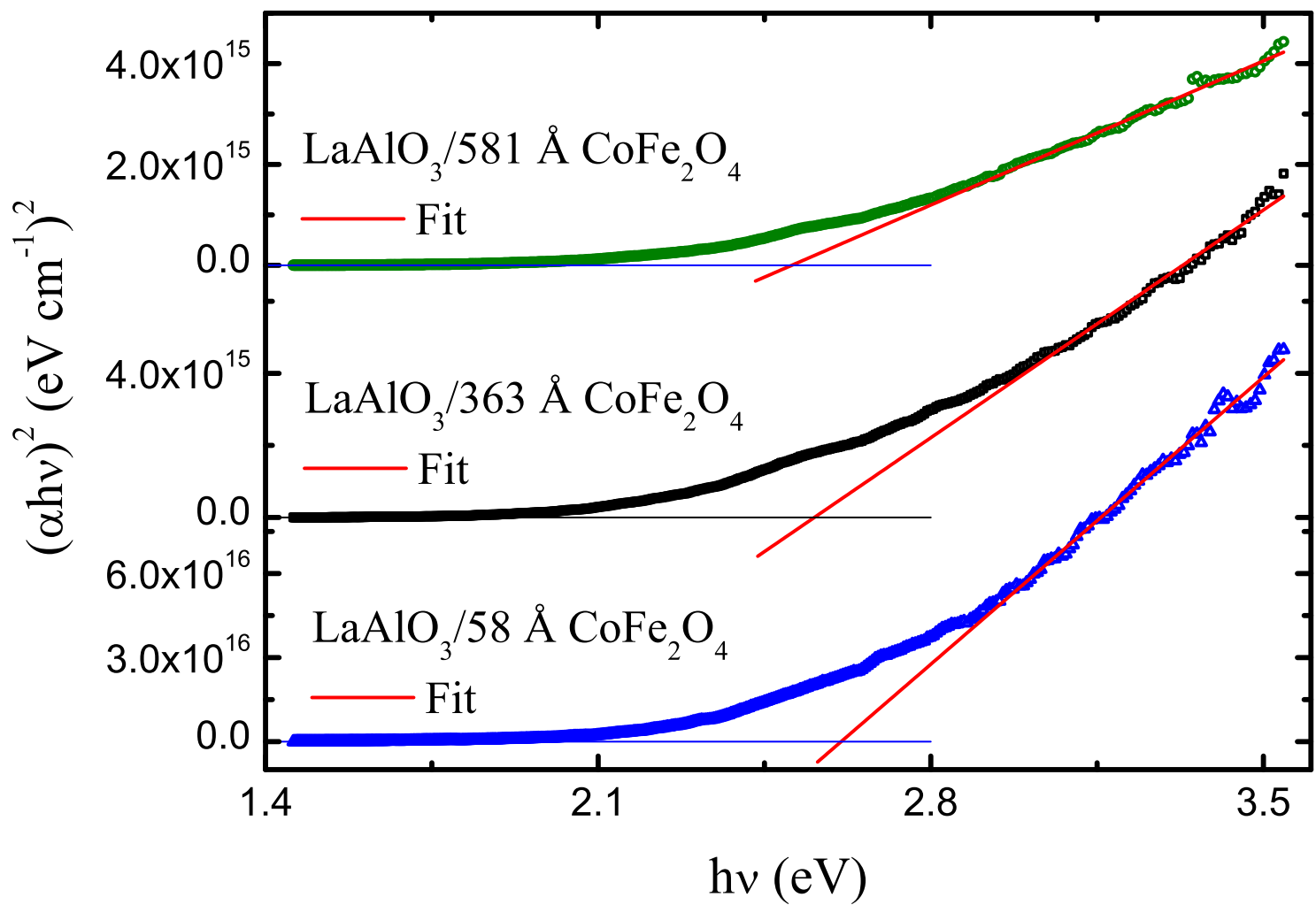